\documentclass[prd,article,twocolumn,preprintnumbers,superscriptaddress,10pt]{revtex4-1}
\usepackage{graphicx,amsfonts,color,comment,amsmath,float}
\usepackage{amssymb}

\usepackage{mathrsfs,amssymb}  
\usepackage{cancel}
\usepackage[normalem]{ulem}

\newcommand{\be}{\begin{equation}}
\newcommand{\ee}{\end{equation}}
\newcommand{\bea}{\begin{eqnarray}}
\newcommand{\eea}{\end{eqnarray}}
\usepackage[dvipsnames]{xcolor}

\newcommand{\scn}{\,\text{sec}}

\usepackage[colorlinks=true]{hyperref}

\definecolor{newgreen}{rgb}{0., 0.56, 0.}

\hypersetup{
    colorlinks=true,
    linkcolor=newgreen,
    filecolor=magenta,      
    urlcolor=Purple,
    citecolor=BrickRed,
}

\begin{document}

\title{Light curves of BSM-induced neutrino echoes in the optically-thin limit}

\author{Ryan Eskenasy}
\affiliation{Department of Physics, The Pennsylvania State University, University Park, Pennsylvania 16802, USA}
\affiliation{Department of Astronomy and Astrophysics, The Pennsylvania State University, University Park, Pennsylvania 16802, USA}
\author{Ali Kheirandish}
\affiliation{Department of Physics, The Pennsylvania State University, University Park, Pennsylvania 16802, USA}
\affiliation{Department of Astronomy and Astrophysics, The Pennsylvania State University, University Park, Pennsylvania 16802, USA}
\affiliation{Center for Multimessenger Astrophysics, Institute for Gravitation and the Cosmos, The Pennsylvania State University, University Park, Pennsylvania 16802, USA}
\author{Kohta Murase}
\affiliation{Department of Physics, The Pennsylvania State University, University Park, Pennsylvania 16802, USA}
\affiliation{Department of Astronomy and Astrophysics, The Pennsylvania State University, University Park, Pennsylvania 16802, USA}
\affiliation{Center for Multimessenger Astrophysics, Institute for Gravitation and the Cosmos, The Pennsylvania State University, University Park, Pennsylvania 16802, USA}
\affiliation{School of Natural Sciences, Institute for Advanced Study, Princeton, New Jersey 08540, USA}
\affiliation{Center for Gravitational Physics and Quantum Information, Yukawa Institute for Theoretical Physics, Kyoto, Kyoto 16802, Japan}

\begin{abstract}
High-energy neutrinos from astrophysical transients serve as a probe of neutrino physics beyond the Standard Model. In particular, nonstandard interaction of neutrinos with the cosmic neutrino background or dark matter (DM) may have imprints on not only their spectra but also the arrival and time-delay distributions. Assuming that the interaction occurs at most once during the neutrino propagation, we provide general analytic formulas for light curves of the neutrino echoes induced by BSM. The formulas can be used for constraining neutrino-neutrino scattering, neutrino-DM scattering, and other scattering processes experienced by relativistic particles. 
\end{abstract}


\maketitle

\section{Introduction}
While ideal cosmic messengers, astrophysical neutrinos offer a unique opportunity to probe for physics beyond the Standard Model (BSM). 
They possess the longest baselines and highest neutrino energies, which empower them to accent the new physics footprints. Observation of the all-sky high-energy cosmic neutrino flux in IceCube \cite{Aartsen:2013bka,Aartsen:2013jdh,Aartsen:2014gkd} has revealed a predominantly extragalactic flux of neutrinos that dominate the neutrino sky in the energy range of TeV to tens of PeV. The presence of any BSM-induced phenomena will introduce signatures in neutrino observables \cite{Arguelles:2019rbn}. Potential features in the spectrum, as well as the arrival direction of cosmic neutrinos, have been exploited to study secret neutrino interactions \cite{Ioka:2014kca,Ibe:2014pja,Ng:2014pca,Blum:2014ewa,Shoemaker:2015qul,Arguelles:2017atb,Araki:2017rex, Bustamante:2020mep,Esteban:2021tub}.
In addition, the precise measurements of the flavor composition of the astrophysical neutrinos would result in stringent constraints on the new physics scenarios \cite{Arguelles:2015dca,Bustamante:2015waa,Shoemaker:2015qul}. 
Yet, another observable that could provide a more powerful tool to study BSM physics is the arrival time of the cosmic neutrinos. The role of this observable in the search for physics beyond the SM is magnified with the recent progress in the identification of high-energy cosmic neutrino sources in time-dependent searches \cite{IceCube:2018cha,IceCube:2018dnn,Amon2018,Stein:2020xhk,Reusch:2021ztx}.

Time-domain multimessenger astrophysics is the primary channel for the identification of the sources of high-energy neutrinos. Transient astrophysical phenomena such as blazar flares \cite{Atoyan:2001ey,Halzen:2005pz,Dermer:2012rg,Dermer:2014vaa,Petropoulou:2016ujj,Halzen:2016uaj,Gao:2016uld}, $\gamma$-ray bursts (GRBs) \cite{Waxman:1997ti,Murase:2005hy,Murase:2008sp,Wang:2008zm,Li:2011ah,He:2012tq,Hummer:2011ms, Kimura:2017kan,Kimura:2018vvz,Biehl:2017qen}, tidal disruption events \cite{Murase:2008zzc,Wang:2011ip,Dai:2016gtz,Senno:2016bso,Lunardini:2016xwi} and supernovae \cite{Meszaros:2001ms,Razzaque:2004yv,Ando:2005xi,Murase:2013ffa,Murase:2009pg,Fang:2015xhg,Fang:2018hjp,Murase:2010cu,Murase:2017pfe} are among the promising sources of high-energy neutrinos. Neutrino source searches are generally more sensitive to transient sources, benefiting from the low background rate. For more details on the potentials of transients in multimessenger astrophysics, see e.g., \cite{Murase:2019tjj}. 

Given that transient sources of astrophysical neutrinos are more likely to be found via multimessenger observations, the arrival time of neutrinos compared to the arrival time of photons and gravitational waves can provide important clues about new physics in the neutrino sector. For high-energy neutrinos, the time delay induced by the neutrino mass is negligible compared to the typical duration of astrophysical transients. 
Hence, any considerable time delay in the arrival of neutrinos at Earth acts as a smoking gun for the presence of new physics either at the source or during their propagation towards the Earth. Murase and Shoemaker (MS19) \cite{Murase:2019xqi} suggested that the time delay induced by neutrino interactions beyond the SM, such as interaction with cosmic neutrino background (C$\nu$B) and dark matter (DM), can be used as powerful probes of the strength of such interactions (see also Ref.~\cite{Koren:2019wwi}). 
Improved sensitivity of the next-generation neutrino detectors such as IceCube-Gen2 will enhance the feasibility of detecting multiple neutrinos from transient sources, which will further strengthen BSM searches via time-domain multimessenger astrophysics. 

New neutrino interactions induced by BSM, a.k.a secret neutrino interactions, naturally arise in a variety of models that attempt to explain the neutrino mass \cite{Chikashige:1980ui,Gelmini:1980re,Blum:2014ewa,Berryman:2022hds} or resolve tensions in cosmological measurements \cite{Aarssen:2012fx,Cherry:2014xra,Loeb:2010gj,Tulin:2012wi,Kaplinghat:2015aga,Tulin:2017ara,Berryman:2022hds}. 
Furthermore, secret neutrino interactions can provide the SM portal to DM \cite{Blennow:2019fhy}, which can facilitate generation of neutrino mass through interaction with the dark sector \cite{Boehm:2006mi,Farzan:2012sa,Escudero:2016tzx,Escudero:2016ksa,Hagedorn:2018spx,Alvey:2019jzx,Patel:2019zky, Baumholzer19, Coito:2022kif}. 
Moreover, secret neutrino interaction can help explain the anomalies in accelerator experiments \cite{Araki:2014ona,Bertuzzo:2018itn,Ballett:2018ynz,Ballett:2019cqp,Ballett:2019pyw, Carpio:2021jhu}.

The magnitude of the time delay, as well as the temporal distribution of the neutrinos arriving at the detector, depend on the number of interactions that high-energy neutrinos undergo during propagation towards the Earth. In the optically thick regime, with a large coupling of interactions or high density of targets, the multiple scattering triggers neutrino cascades \cite{Ioka:2014kca,Ng:2014pca,Arguelles:2017atb}. Hence, the majority of neutrinos would emerge at lower energies. Conversely, in the optically thin limit, single neutrino interaction is expected, and therefore neutrinos may arrive at similar energies as their site of production.

In this study, we focus on secret neutrino interaction in the optically thin limit and provide the general formulation to calculate scattered neutrino signals. 
The main advantage of considering this limit is that it can easily accommodate the present constraints on the BSM interactions from other channels of observation and/or experiments and the analytical results can be confronted with those of numerical simulations. Additionally, neutrinos will arrive at similar energies they had at their site of production. Therefore, the time-delay profile does not depend on the initial fluence and spectrum \cite{Murase:2019xqi}. We explore the time-delay structure for high-energy neutrinos and present the ``light curves'' for the neutrino emission in the optically thin limit. For this purpose, we focus on two specific BSM-induced neutrino interactions: high-energy neutrino interaction with the C$\nu$B and neutrino interaction with DM, but the formalism can be applied to arbitrary single-scattering process experienced by relativistic particles. 

Throughout this paper, we use natural units and set $\hbar = c= 1$. This paper is organized as the following: in Sec. \ref{sec:form}, we discuss the characteristics of small angle scatterings and present the formulation of the temporal information for the delay in the arrival time of neutrinos. In Sec. \ref{sec:lc}, we provide examples of light curves for neutrino emission for secret neutrino interactions.

\section{Analytical Light Curves of Neutrino Echoes}\label{sec:form}
\begin{figure}[tb]
    \centering
    \includegraphics[width=1\columnwidth]{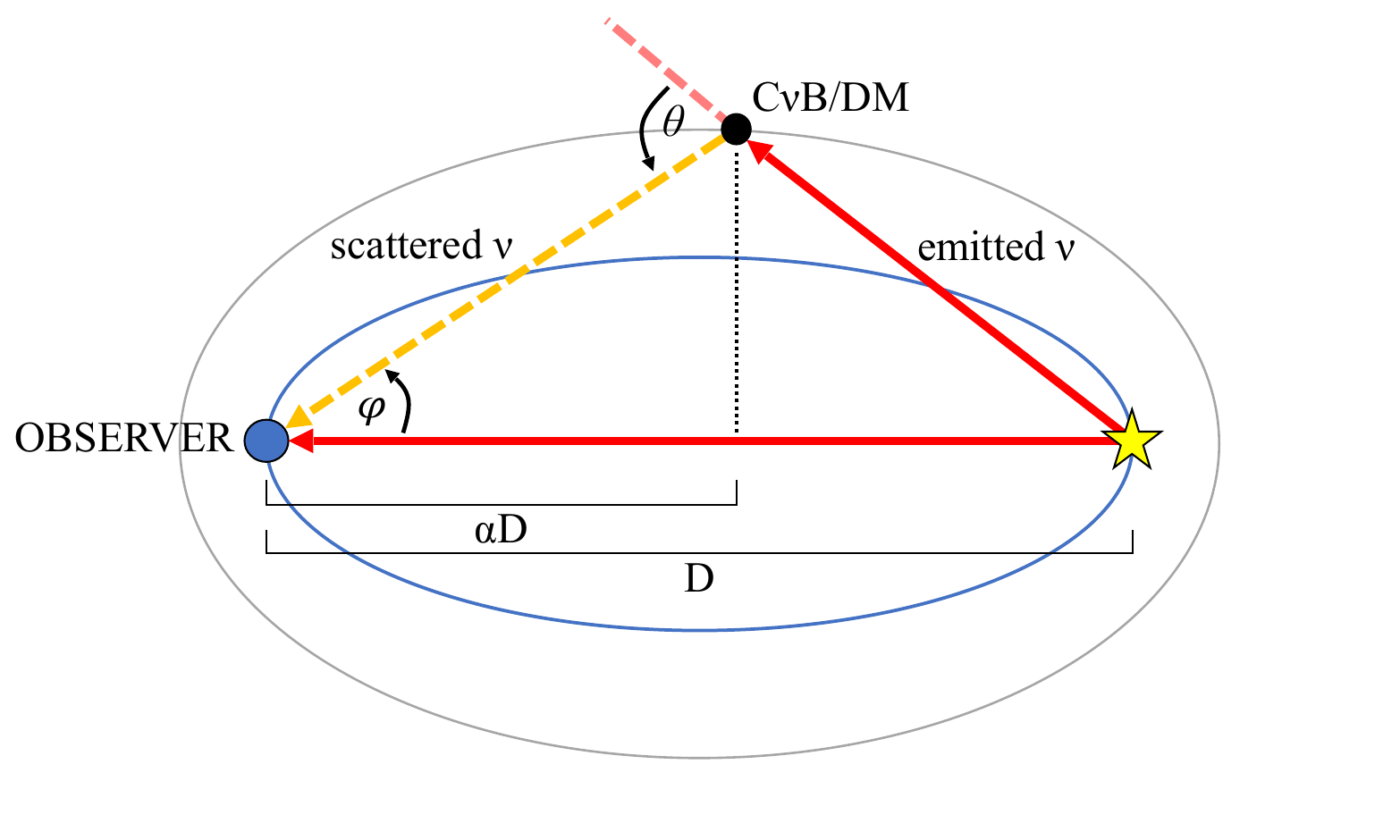}
    \caption{Schematic illustration of neutrino echoes induced by BSM neutrino interactions with the C$\nu$B or DM.}
    \label{fig:schematic}
\end{figure}

Time delay induced by high-energy particle interactions in the optically thin regime may result from small scattering angles with respect to the original direction of the particle. 
In the astrophysical context, early studies studied such small-angle scatterings to explain the tails of X-ray bursts \cite{1978ApJ...222..456A}. The key parameter defining the regime of the scattering is the optical depth for nonstandard neutrino interaction. The geometrical setup is somewhat analogous to $\gamma$-ray ``pair echoes'' proposed as a probe of intergalactic magnetic fields~\cite{Plaga:1995ins,Ichiki:2007nd,Murase:2008pe,Takahashi:2008pc,Murase:2008yy,Murase:2009ah}, although the underlying interaction processes are completely different. 

The cosmic distances that high-energy neutrinos travel can provide the required mean free path, i.e., sufficient optical depth, for their interaction with the C$\nu$B or DM. The generic framework for such interaction is illustrated in Fig.~\ref{fig:schematic}. The locale of scattering sites giving a fixed time delay is an ellipsoid of revolution with foci at the source and the observer. An interaction alters the directions of the neutrinos and increases the distance they need to travel towards the observer. The deviating angle for the high-energy neutrino depends on the characteristics of the interaction, which is reflected by the differential cross section.

The differential cross section for the two-body scattering $1+2\rightarrow3+4$ in the rest frame of $2$ is given by:
\be 
\frac{d\sigma}{d\Omega}=\frac{1}{64 \pi^{2}p_1m_1} \frac{{|p_{3}|}^2}{|p_{3}|(E_1+m_2)-E_3|p_1|\cos\theta} |\mathcal{M}|^{2},
\ee
where $|\mathcal{M}|^{2}$ is the spin-averaged matrix elements, $p$ and $E$ are the particle's momentum and energy, respectively, and $\theta$ is the scattering angle.

For the differential cross section ($d\sigma/d\Omega$), the average scattering angle, $\theta$, is evaluated via 
\be
\langle(1-\cos\theta)\rangle=\frac{1}{\sigma}\int d\Omega \, (1-\cos\theta) \left(\frac{d\sigma}{d\Omega}\right), \label{eq-aveang}
\ee 
where $\sigma$ is the total cross section. One could average over $\theta^2$ in this equation. However, in the optically thin limit, the scattering angle is small and the left hand side of Eq.~\ref{eq-aveang} can be replaced by $\langle \theta^2/2 \rangle$.

The probability of a neutrino scattering off, echoing, is determined by the optical depth $\tau_\nu \approx n \sigma_\nu D$, where $n$ is the number density of background particles and $D$ is the distance from their origin. In the optical-thin limit, neutrinos would experience at most one scattering, provided that the distance they travel is longer than the mean free path for the nonstandard interaction.

The additional distance that neutrinos would travel depends on the scattering angle distribution. In the next section, we present the geometrical formulation for this scattering angle and discuss the probability distribution of the time-delayed neutrinos.

\subsection{General Formulation}
The geometrical configuration for neutrino echoes is described by an ellipsoid with foci located at the position of the observer and source (see Fig.~\ref{fig:schematic}). Introducing the distance between the source/observer and scattering point, $A/B$, the time delay ($t$) can be expressed as 
\begin{equation}
t=A+B-D
\end{equation}
where the speed of neutrinos is set to the speed of light, $c=1$.   
 
Using the law of sines, the fractional distance where the scattering takes place to the observer is given by
\begin{equation}
\alpha= \frac{\kappa}{2} \bigg( \frac{\kappa+2}{\scn\, \varphi (\kappa+1)-1} \bigg).
\label{eq-alpha}
\end{equation}
Here $\kappa\equiv t/D$ and $\varphi$ is the neutrino's arrival angle with respect to the source direction, and we note $0
\leqq \alpha\leqq1$.

Using the law of cosines, $D^2=A^2+B^2+2AB\cos\theta$, the scattering angle is expressed as 
\begin{widetext}
\begin{equation}
\cos\theta=  \frac{\sqrt{1-\sin^2\varphi} - [\kappa(1+\kappa/2)-\kappa\sqrt{1-\sin^2\varphi}+\kappa^2{(1+\kappa/2)}^2\mathrm{cosec}^2\varphi]}{{(1+\kappa)}^2+\kappa^2{(1+\kappa/2)}^2\mathrm{cosec}^2\varphi}.
\label{eq-theta1}
\end{equation}
\begin{equation}
\sin\theta= \frac{(1+\kappa)\sin\varphi + [\kappa(1+\kappa)(1+\kappa/2)+\kappa\sqrt{1-\sin^2\varphi}+\kappa^2\sqrt{1-\sin^2\varphi}/2]\mathrm{cosec}\varphi}{{(1+\kappa)}^2+\kappa^2{(1+\kappa/2)}^2\mathrm{cosec}^2\varphi}.
\label{eq-theta2}
\end{equation}
\end{widetext}
Form the geometrical constraint, we have the lower-limit of the scattering angle as $\theta_{\rm min}=2\mathrm{arccos}[{(1+\kappa)}^{-1}]$. The maximum scattering angle is $\theta_{\rm max}=\pi$.

Eq.~\ref{eq-alpha} and \ref{eq-theta2} are then used to build the probability distribution of delayed neutrinos in the optically thin limit. The probability distribution function is 
\begin{equation}
P(t,\varphi; D) = J(t,\varphi)\frac{1}{\sigma_\nu}\left(\frac{d\sigma_\nu}{d\cos\theta}\right),
\label{eq-general}
\end{equation}
where
\begin{equation}
J(t,\varphi)=\left|\frac{\partial\alpha}{\partial{t}}\frac{\partial\cos\theta}{\partial\varphi} - \frac{\partial\alpha}{\partial\varphi}\frac{\partial\cos\theta}{\partial{t}}\right| 
\end{equation}
is the determinant of the Jacobian matrix, and $\sigma_\nu$ is the cross section for BSM neutrino interactions. 

For practical purposes, it is worthwhile to mention how the flux of scattered neutrinos, $\frac{dN_{\rm echo}}{dEdt}$ can be calculated. 
Noting that $\theta$ is a function of the incident particle energy $E'$ and the scattered particle energy $E$, one can change the variable from $\varphi$ to $E'$ for a given $E$. Then, we have   
\begin{equation}
\frac{dN_{\rm echo}}{dEdt} = \int dE' \frac{dN_{\rm source}}{dE'}
{\left|\frac{d\cos\theta}{d\varphi}\right|}^{-1}J
\left(\frac{d\sigma_\nu}{dE}\right) n D
,
\label{eq-general}
\end{equation}
where $\frac{dN_{\rm source}}{dE'}$ is the spectrum of neutrinos at the source. See also Ref.~\cite{Carpio:2022sml}.

The general formulas derived above can be applied to even large-angle scatterings that are often difficult to numerically simulate. Examples include neutrino-DM scattering that could occur in the Galactic halo.

\subsection{Small-Angle Scattering Limit}
Eq.~(\ref{eq-general}) gives the general formula for light curves of neutrino echoes even if the scattering angle is large. However, for astrophysical sources, $\kappa \ll 1$ is almost always satisfied. Also, for most of applications to high-energy cosmic neutrinos, the scattering angle is expected to be small enough, in which the time delay is not too long to be observable. 
Thus, it is useful to consider the small scattering angle limit, i.e., $\varphi^2 \ll 1$ and $\kappa \ll 1$.  In this limit, Eq.~(\ref{eq-alpha}) leads to
\begin{equation}
\alpha \approx\frac{2\kappa}{2\kappa+\varphi^2},
\label{eq-alphaapprox}
\end{equation}
and Eq.~(\ref{eq-theta1}) becomes
\begin{equation}
\theta \approx \varphi + \frac{2\kappa}{\varphi},
\end{equation}
in the leading order. 

Then, for a given $\varphi^2$, the probability distribution function for the arrival time of delayed neutrinos is expressed as 
\be
P(t,\varphi; D)\approx \frac{1}{t+(D\varphi^2/2)}\frac{1}{\sigma_\nu}{\left(\frac{d\sigma_\nu}{d\theta}\right)}_{\theta=\varphi+2t/(D\varphi)}.
\label{eq:analy}
\ee
which reproduces the formulas presented by MS19~\cite{Murase:2019tjj} and Ref.~\cite{1978ApJ...222..456A}.
Here we remark that only one scattering matters and the time-delay distribution reflects the differential cross section of the nonstandard neutrino interaction that is generally inelastic. 

The final time-delay distribution is calculated by
\be
P(t)\approx \int \frac{d\varphi}{t+(D\varphi^2/2)}\frac{1}{\sigma_\nu}{\left(\frac{d\sigma_\nu}{d\theta}\right)}_{\theta=\varphi+2t/(D\varphi)},
\label{eq:analy2}
\ee
where the geometric constraint should be imposed for $\varphi$ to satisfy $\theta_{\rm min}\leqq\theta\approx \varphi+2t/(D\varphi)\leqq\theta_{\rm max}$, where $\theta_{\rm min}=2\sqrt{2t/D}$ and $\theta_{\rm max}=\pi$. 

In the next section, we study the time-delay profile and present the light curve for the two motivated BSM scenarios.

\section{Applications}\label{sec:lc}
Here, as demonstrative examples, we consider three BSM models that could induce time-delay signatures in the arrival time of high-energy cosmic neutrinos, assuming the optically thin limit. The first two interactions that we consider are the high-energy neutrinos bouncing off the C$\nu$B. 
We consider the $s$-channel interactions for such neutrino-neutrino interactions via both scalar and vector mediators. Moreover, we will consider DM-neutrino interaction. For this purpose, we assume fermionic DM interacting with high-energy neutrinos via a scalar and vector mediator. 

\begin{figure}[t]
    \centering
    \includegraphics[width=\columnwidth]{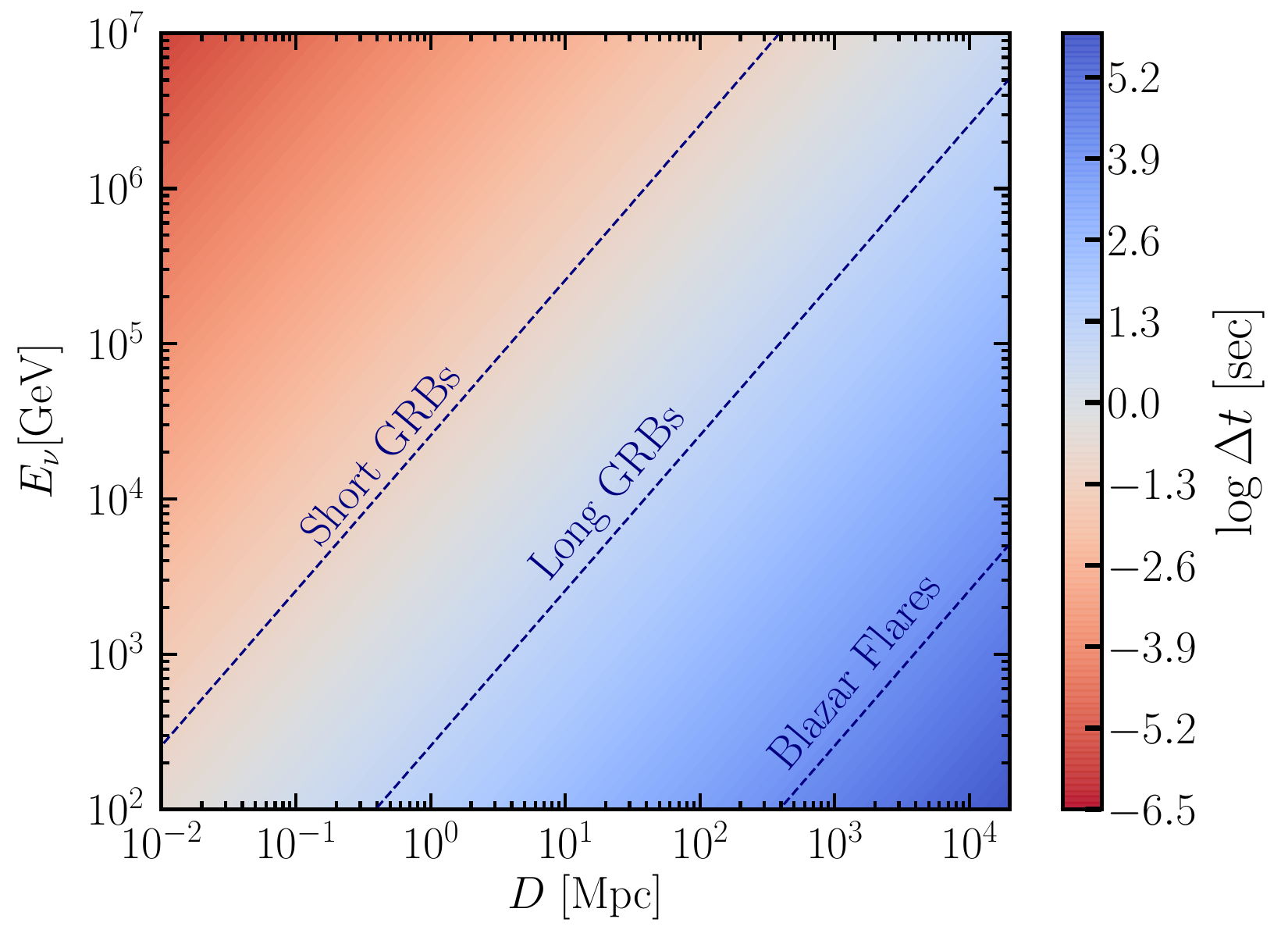}
    \caption{Expected time delay induced by neutrino-neutrino interactions ($s$-channel, scalar mediator) for different combination of neutrino energy and source distances. Dashed lines show the typical timescale for example of transient sources: short GRBs ($10^{-1}$ s), long GRBs ($10^{2}$ s), and blazar flares ($10^{5}$ s).}
    \label{fig:contour}
\end{figure}

\subsection{Neutrino self-interaction}
Neutrinos can self-interact in the SM by exchanging the $Z$ boson. However, because of the large mass of $Z$, such an interaction with the C$\nu$B would only leave a trace in the cosmic neutrino's observable when the cosmic neutrino's energy exceeds $\sim10^{22}$~eV \cite{Weiler:1982qy}. On the other hand, considerable neutrino self-interaction is expected in a variety of BSM scenarios \cite{BialynickaBirula:1964zz,Bardin:1970wq}. Secret neutrino self-interactions can generate finite neutrino masses ~\cite{Chikashige:1980ui,Gelmini:1980re,Blum:2014ewa}. As such, they are regarded as one of the popular, well-motivated, BSM scenarios in the neutrino sector. In addition, the presence of secret neutrino interaction could alleviate the cosmological anomalies, such as the discrepancy in the reported value for the Hubble constant \cite{Escudero:2019gzq,Kreisch:2019yzn,Barenboim:2019tux}, and 
small-scale structure problems ~\cite{Aarssen:2012fx,Cherry:2014xra}.

In this work, we focus on neutrino self-interactions mediated via a scalar or vector mediator. The effective Lagrangian for such interactions would be ${\mathcal L}\supset g_{\nu}\bar{\nu}\nu\phi$ and ${\mathcal L}\supset g_{\nu}\bar{\nu}(\gamma^\mu V_\mu)\nu$ for scalar and vector mediators, respectively. In the scalar mediator case, the squared matrix element for neutrino-neutrino scatterings is,
\bea
|\mathcal{M}|^{2}_{\nu\bar{\nu}\rightarrow\nu\bar{\nu}} &= &g_{\nu}^{4} \biggl[\frac{s^2}{(s-m_{
\phi}^{2})^{2} + \Gamma_{\phi}^{2} m_{\phi}^{2}}\biggr]
\eea
where $g_\nu$ is the coupling and $m_\phi$ is the mediator mass. Note that the Majorana case can be similarly treated by using $n_\nu=0.5n_{\nu+\bar{\nu}}$ \cite{Blum:2014ewa}.  

Here, examine the $s$-channel interaction with a scalar mediator, in which the amplitude is spin independent. This particular case includes a resonance interaction which occurs within the range of cosmic neutrino flux (see e.g., Refs.~\cite{Ioka:2014kca,Ng:2014pca,Ibe:2014pja,Araki:2014ona,Blum:2014ewa,Cherry:2014xra,DiFranzo:2015qea,Araki:2015mya,Shoemaker:2015qul,Kelly:2018tyg,Barenboim:2019tux}). For the $s$-channel interaction, we have
\begin{equation}
\frac{1}{\sigma_\nu}\frac{d\sigma_\nu}{d\cos \theta} = \frac{E_\nu m_\nu}{(m_\nu+(1-\cos \theta)E_\nu)^2}
\end{equation}

\begin{figure*}[t]
    \centering
    \includegraphics[width=1\columnwidth]{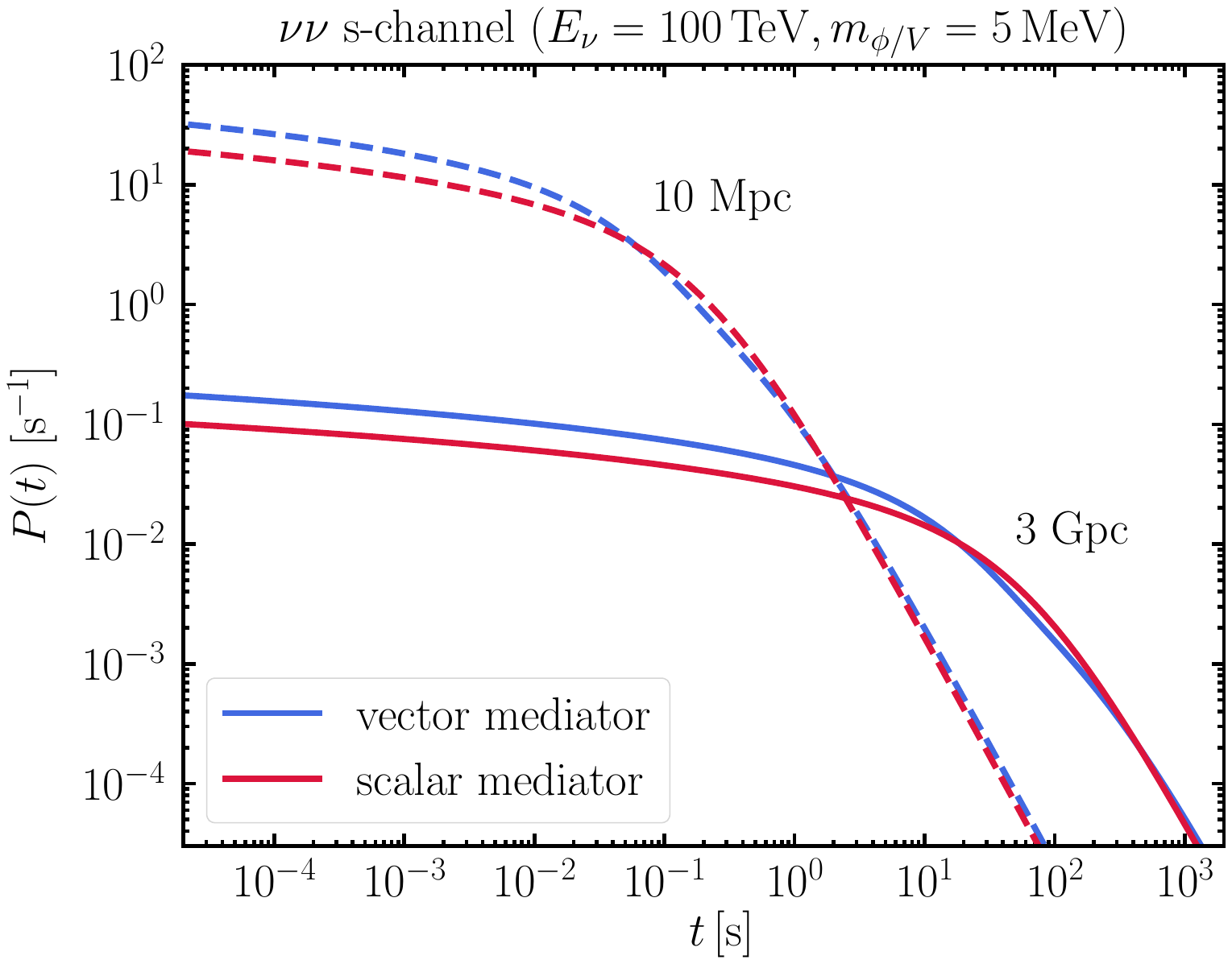}
    \includegraphics[width=1\columnwidth]{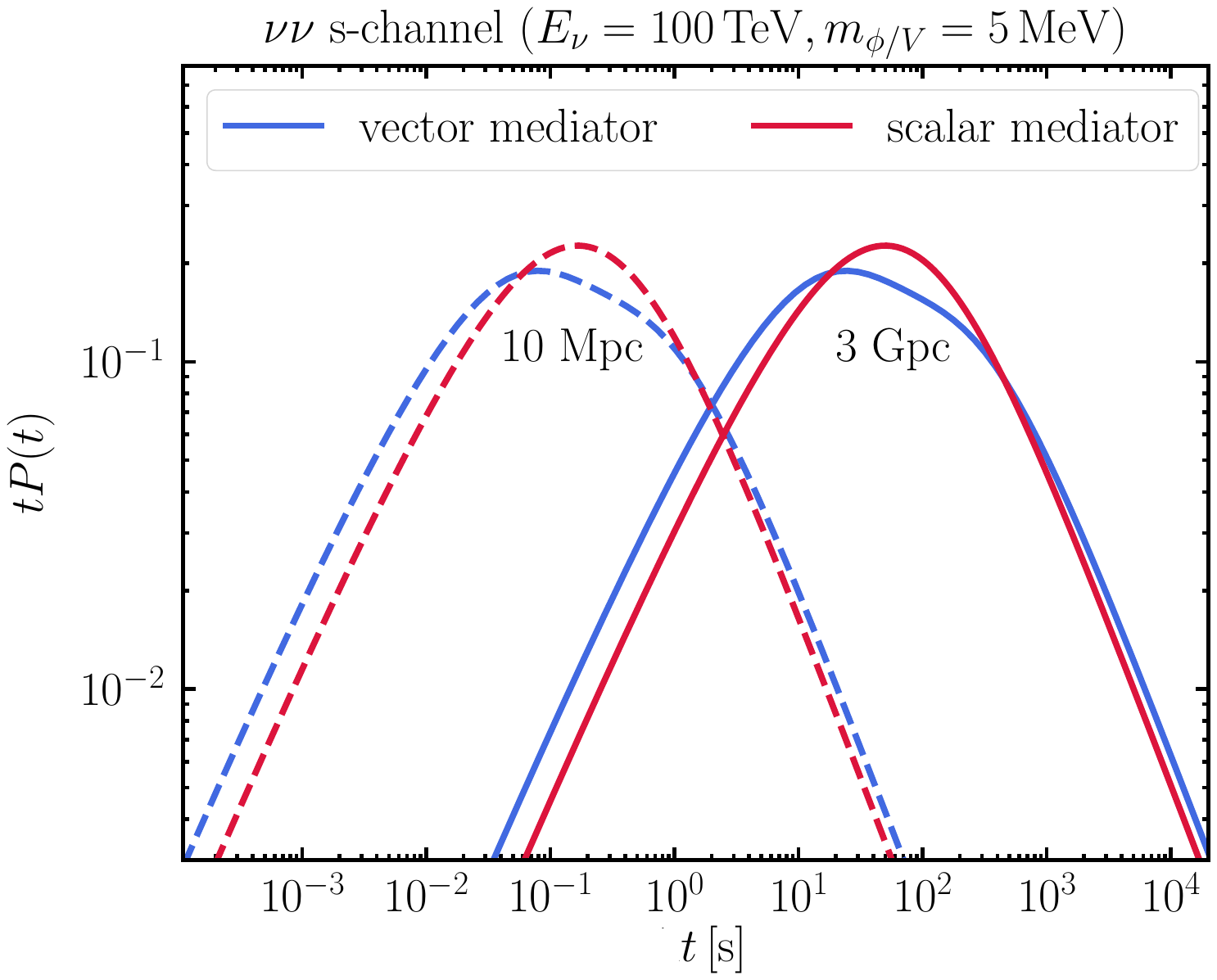}
    \caption{Left panel: $P(t)$ for 100 TeV neutrinos originating from a 10 Mpc (dashed line) and a 3 Gpc (solid line) source. The $\nu\nu$ interaction is shown via a scalar (red line) and vector (blue line) mediator. 
    Right panel: Similar to the left panel, but for $tP(t)$. The characteristic time delay corresponds to the time at the peak of $tP(t)$. 
    }
    \label{fig:nunu-pt}
\end{figure*}

In Fig.~\ref{fig:contour}, we showcase the expected time delay for neutrino self-interaction compared to the typical timescale for examples of transient astrophysical phenomena. Using Eq.~(\ref{eq:analy}), we present the temporal profile for the time-delay neutrino signal in the left panel of Fig.~\ref{fig:nunu-pt} for neutrino self-interaction. We show the profiles for both scalar and vector mediator $s$-channel interactions. The light curve for echoed neutrinos are presented in the right panel. The distribution around the peak of $tP(t)$ differs due to the mediator spin. 
We should note that in the neutrino self interaction also receives contributions from $t-$ and $u-$channels. The $t-$channel, in particular, has a different dependence on the scattering angle. Here, we focused on the $s-$channel cross section for illustration, specially since it involves a resonance. One can follow similar formalism for the $t$-channel. For additional details on $t-$channel cross section for neutrino self-interactions, see Ref.~\cite{Blum:2014ewa}.

\begin{figure}[t]
    \centering
    \includegraphics[width=1\columnwidth]{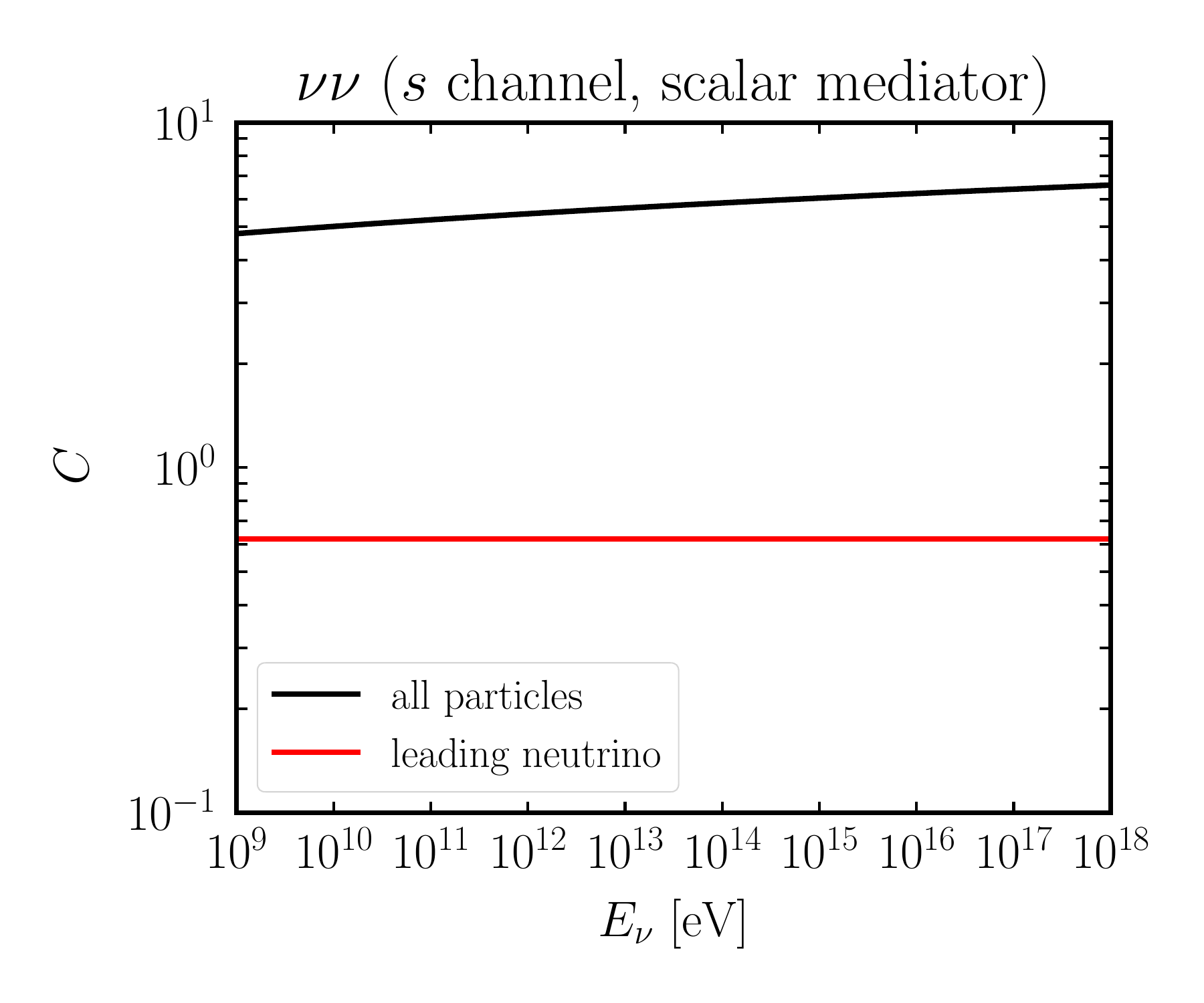}
    \caption{$C$-factors for neutrino self-interaction ($s$-channel) via a scalar mediator. The black line corresponds to the $C$-factor for all particles, and the red line shows the $C$-factor when only the leading neutrinos are considered.}
    \label{fig:cfactor}
\end{figure}

It is useful to give an estimate on the characteristic time delay, which is \cite{Murase:2019xqi}
\begin{equation}
\Delta t \approx\frac{1}{2}\frac{\langle\theta^2\rangle}{4} D\simeq77~{\rm s}~\left(\frac{D}{3~\rm Gpc}\right)C^2{\left(\frac{m_\nu}{0.1~{\rm eV}}\right)}{\left(\frac{0.1~{\rm PeV}}{E_\nu}\right)}.
\label{delay2}
\end{equation}
where $m_\nu$ is the neutrino mass and $C\equiv \sqrt{\langle\theta^2\rangle}E_\nu/\sqrt{s}$ is a factor that depends on particle physics models. Throughout this work, we assume $m_\nu=0.1$ eV.

In the single scattering case, the average scattering angle of leading particles would be the most important \cite{Murase:2019xqi}. To take into account the leading particles contribution only, by using the Lorentz boost, we obtain
\begin{equation}
\langle 1-\cos \theta \rangle_{\rm lead}=\int_0^1 d\cos \theta^\prime \frac{(1-\beta)(1-\cos \theta^\prime)}{1+\beta \cos \theta^\prime},
\end{equation}
where $\beta=\sqrt{1-1/\gamma^2}$ and $\gamma$ is the boost factor $E_\nu/\sqrt{s}$. Fig.~\ref{fig:cfactor} shows the value of $C$ when leading and all neutrinos are considered. Note that $\theta^\prime$ is the direction of neutrinos in the center-of-momentum frame. We reproduce the result on $\langle \theta^2 \rangle$ by MS19, which corresponds to $C\simeq0.62$ for leading neutrinos. 
For all particles, by implementing the $\nu\nu$ differential cross section in Eq.~\ref{eq-aveang}, we use
\begin{equation}
\langle 1-\cos \theta \rangle_{\rm all} =  \frac{m_\nu}{E_\nu}\bigg( 1+\ln ( 1+\frac{2E_\nu}{m_\nu}) \bigg). 
\end{equation}
The characteristic time delay of numerical light curves (i.e., the peak time of $tP(t)$) is in good agreement with Eq.~(\ref{delay2}) with $C$ for leading neutrinos.  

\subsection{Dark matter-neutrino interaction}
Now, we consider BSM interaction of neutrinos with DM. An ongoing decay or annihilation of DM to SM particles implies possible interaction of SM and DM. 
DM-neutrino interaction is especially attractive for light DM models, where annihilation into heavier products is kinematically forbidden, and appears naturally in some models that DM is the sterile neutrino  \cite{Drewes:2016upu}.  Study of secret neutrino interaction with DM has become more attractive with the observation of high-energy cosmic neutrinos as the higher energies would manifest the BSM signatures, where constraints have been obtained using high-energy astrophysical neutrinos~\cite{Farzan:2014gza,Davis:2015rza,Cherry:2016jol,Arguelles:2017atb,Kelly:2018tyg,Farzan:2018pnk,Pandey:2018wvh,Choi:2019ixb,Capozzi:2018bps,Murase:2019xqi}.  In addition, the multimessenger identification of high-energy neutrinos from flaring blazar, TXS 0506+056, has prompted new limits on boosted DM \cite{Wang:2021jic, Granelli:2022ysi}, and our formalism can also be applied to boosted DM by considering the angular distribution of scattered DM particles.
 
In this study, we focus on two scenarios for $t$-channel DM-neutrino interaction, mediated by scalar and vector mediators. In a model in which a scalar couples to neutrinos as well as fermionic DM, $\mathcal{L} \supset g_{\nu} \bar{\nu}\nu\phi + g_{X} \bar{X} X\phi$, only a $t$-channel diagram contributes to the DM-neutrino scattering. This squared matrix element is
\be 
|\mathcal{M}|^{2}_{X\nu\rightarrow X\nu} = \frac{2 g_{\nu}^{2} g_{X}^{2}}{(t-m_{\phi}^{2})^{2}} \left[( 4m_{\nu}^{2} - t)( 4m_{X}^{2} - t) \right].
\ee

In a vector mediator model with a vector mass $m_{V}$ and Dirac fermionic DM, we consider $\mathcal{L}_{V} \supset g_{\nu} V_{\mu}  \bar{\nu}\gamma^{\mu} \nu + g_{X} V_{\mu}  \bar{X} \gamma^{\mu}X$. The squared matrix element for DM-neutrino scatterings is
\bea
|\mathcal{M}|^{2}_{X\nu \rightarrow X\nu} &=& \frac{2 g_{\nu}^{2} g_{X}^{2}}{(t-m_{V}^{2})^{2}} \nonumber\\
 &\times& \left[ s^2+u^2+2m_X^4-2m_X^2(s-t+u)
 \right],\nonumber\\
\eea
which is also used in the previous work \cite{Murase:2019xqi}. Using $s+t+u = 2m_{X}^{2}$, one can eliminate $u$ in the above so that the expressions are only in terms of $s$ and $t$.  Note as well that in the limit of small momentum transfer ($t=0$) we recover the expression used in Ref.~\cite{Aarssen:2012fx} for DM-neutrino bounds from the late decoupling. 

\begin{figure}[t]
    \centering
    \includegraphics[width=\columnwidth]{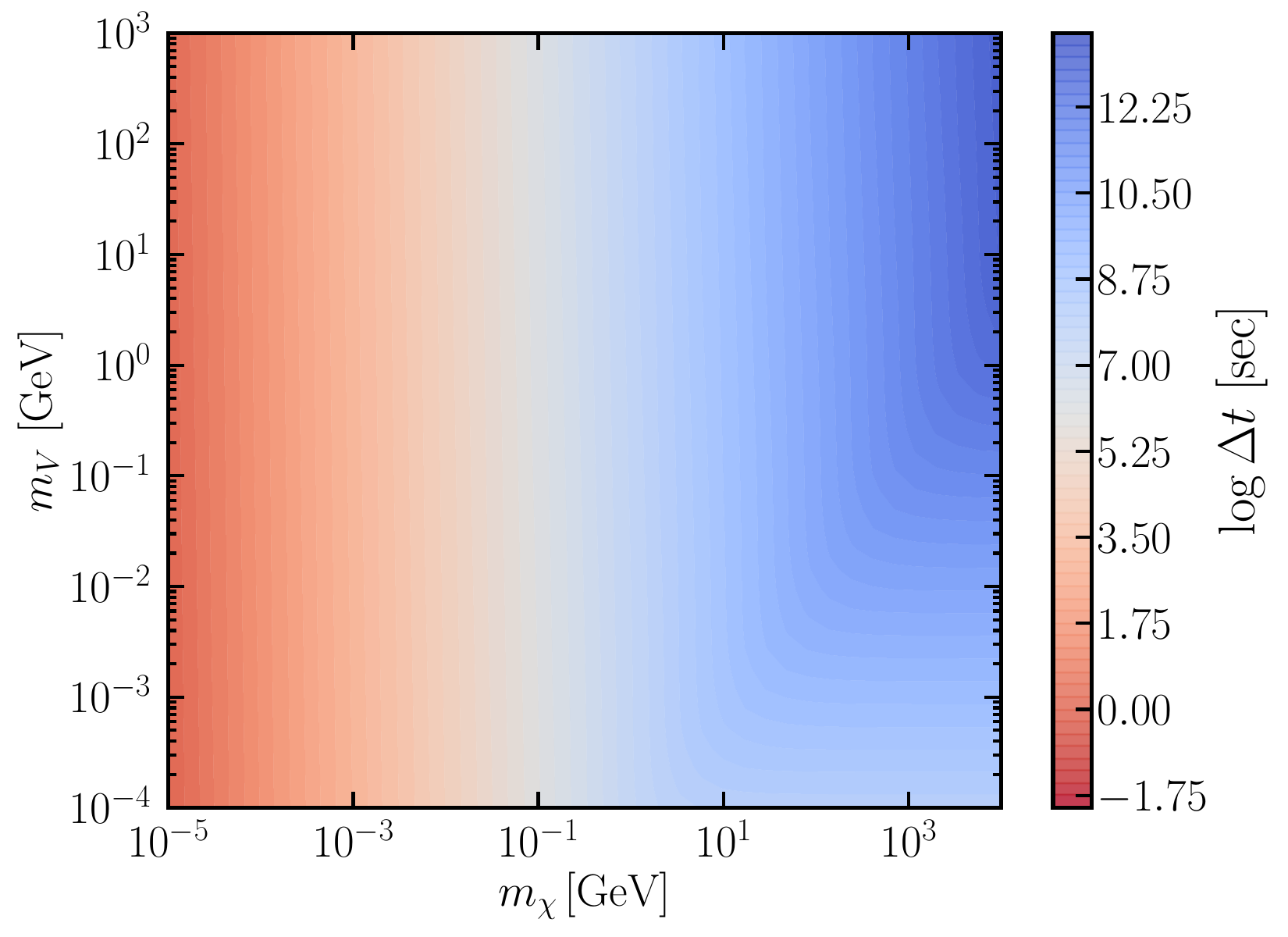}
    \caption{Time delays induced by DM-neutrino interaction via a vector mediator ($t$-channel) for 100 TeV neutrinos emitted by a source at 3~Gpc.}
    \label{fig:DM-contour}
\end{figure}

\begin{figure*}
    \centering
    \includegraphics[width=1\columnwidth]{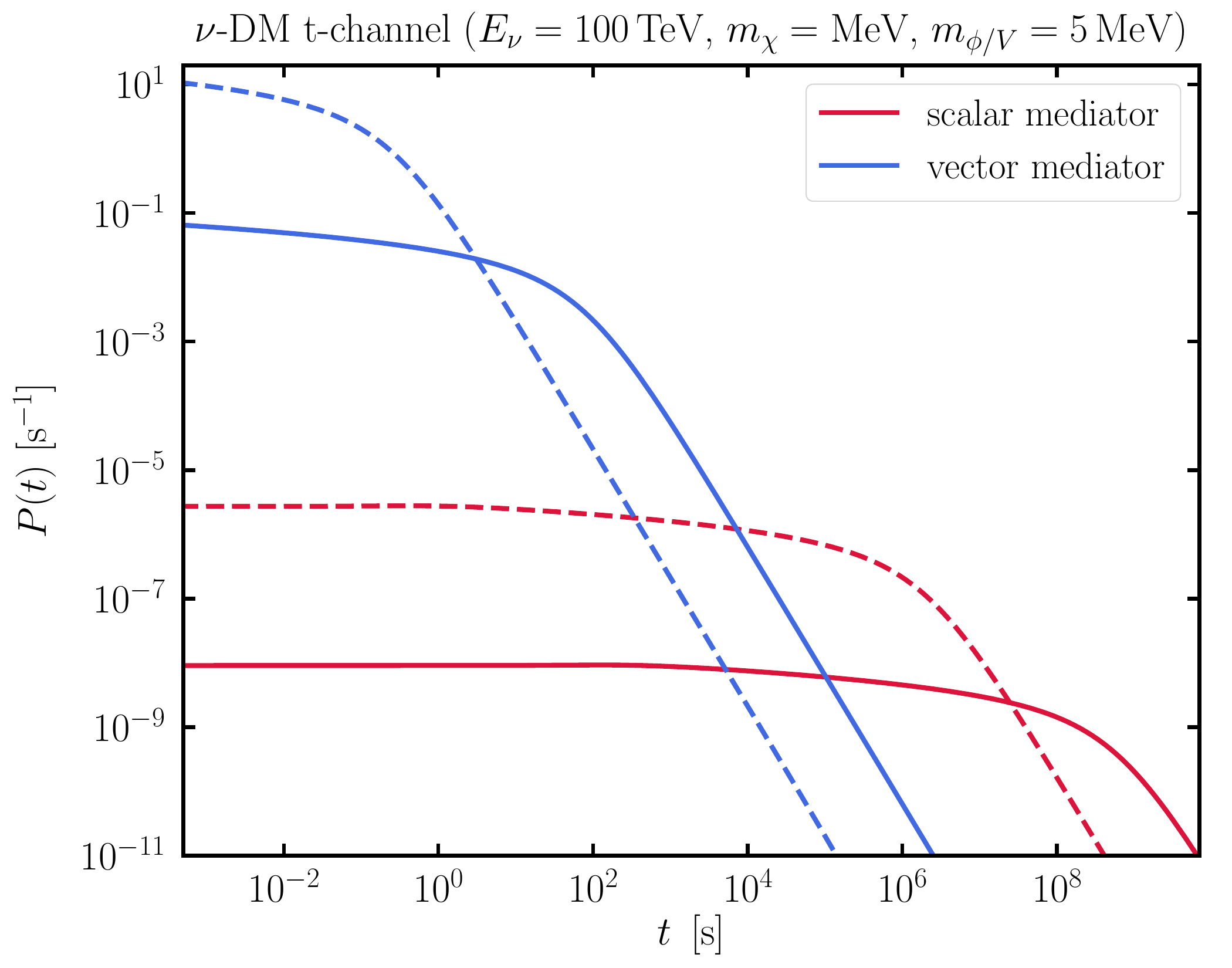}
    \includegraphics[width=1\columnwidth]{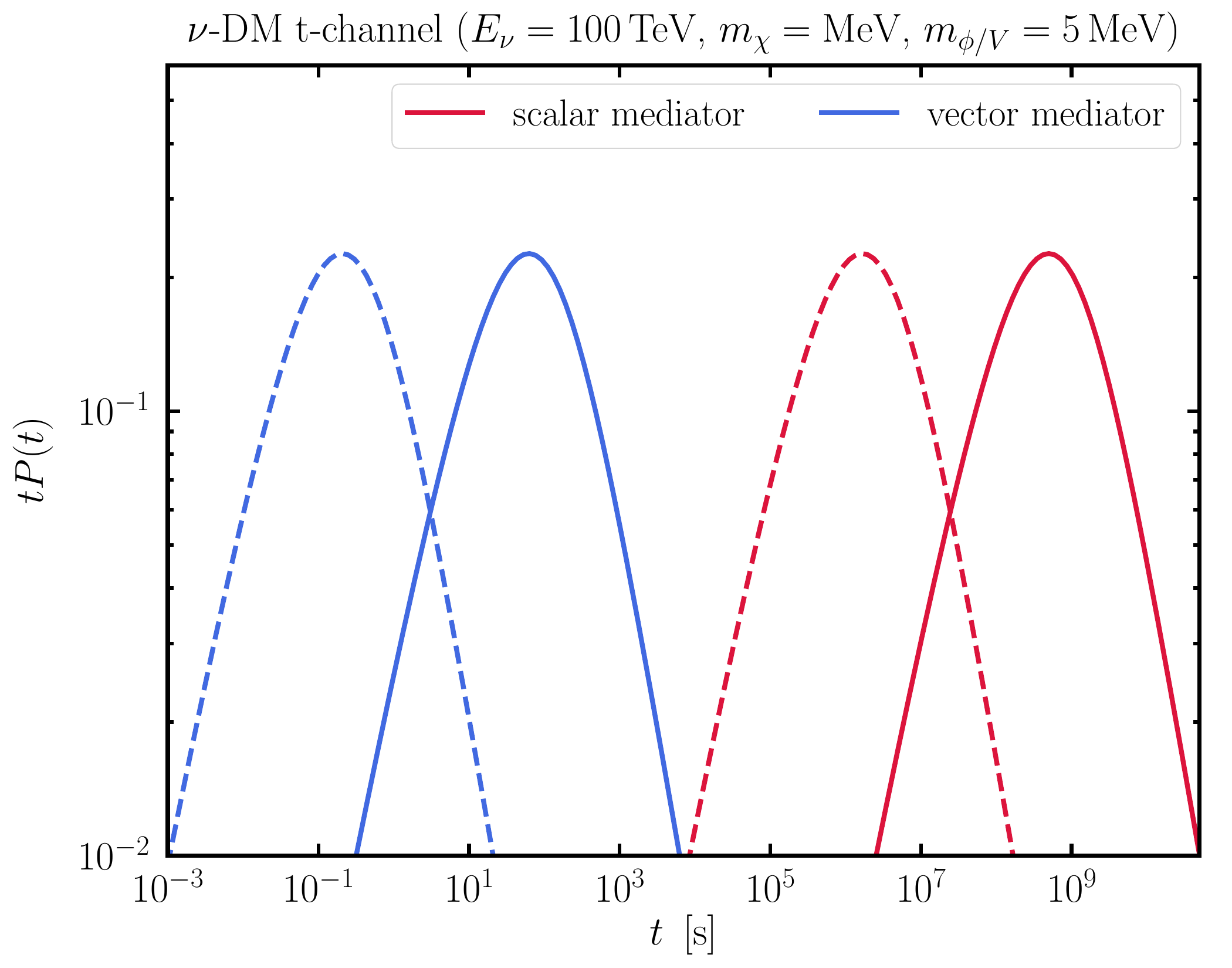}
    \caption{Left panel: $P(t)$ for 100 TeV neutrinos originating from a 10 Mpc (dashed line) and a 3 Gpc (solid line) source. The $\nu$-DM interaction is shown via a scalar (red line) and vector (blue line) mediator. 
    Right panel: Similar to the left panel, but for $tP(t)$. The characteristic time delay corresponds to the time at the peak of $tP(t)$.}
    \label{fig:ptDM}
\end{figure*}

Figure~\ref{fig:DM-contour} shows the average time delay for neutrino interaction via vector mediator for different combinations of DM ($m_\chi$) and the vector mediator ($m_{\rm V})$. A specific feature for this interaction is that for $m_\chi < \rm \, GeV$, the DM-neutrino cross section for this scenario becomes almost independent of the mediator mass. Similar to self neutrino interactions, we present the temporal distribution and the light curves for the neutrino-DM interactions in Fig.~\ref{fig:ptDM}. While for self neutrino-interaction, the peaks of the light curves were close for scalar and vector mediator, we find that neutrino-DM interaction via the vector mediator scenario leads to a shorter time delay in the arrival of high-energy neutrinos compared to interaction via a scalar mediator.

\section{Summary and Discussion}
Identifying the sources of high-energy neutrinos will offer unique opportunities to probe BSM physics in the neutrino sector. In this work, we presented light curves of BSM-induced neutrino echoes. We provided an analytical formulation of the time-delay distribution for neutrinos scattering off the background. This study is useful for more detailed studies using the Monte Carlo simulations~\cite{Carpio:2022lqk}. 

We focused on neutrino-neutrino and DM-neutrino interactions, which are highly motivated and can ease tensions present in cosmological and accelerator measurements. Thanks to their high energies and long baselines, observation of high-energy cosmic neutrinos opened a new front for studying such interactions, which outperforms cosmological measurements in part of the parameter space. The majority of previous studies focused on the features expected in the energy spectrum or arrival direction of high-energy neutrinos, see e.g., Refs.~\cite{Ioka:2014kca,Ng:2014pca,Blum:2014ewa,Davis:2015rza,Shoemaker:2015qul,Cherry:2016jol,Arguelles:2017atb,Kelly:2018tyg,Pandey:2018wvh,Choi:2019ixb,Capozzi:2018bps}.
Recent progress in the identification of the origin of high-energy neutrinos with multimessenger observations has provided an additional tool for probing new physics in the neutrino sector. Incorporating the arrival time, and investigating the deviation from the arrival of other electromagnetic radiation, or gravitational waves not only opens a new avenue in studying new physics but also provides a complementary and competing probe of the new physics with high-energy cosmic neutrinos.

As shown in MS19 \cite{Murase:2019xqi}, new physics searches with time-domain multimessenger neutrino astrophysics will provide a competitive probe to the previous searches with astrophysical neutrinos and cosmological measurements. Recent additional coincidences between high-energy neutrinos and flaring sources demonstrate the feasibility and power for new physics studies with neutrino echoes. The enhanced sensitivity of the next generation of neutrino telescopes will help us identify multiple neutrinos from a transient source and study their arrival time profile.  
An important point is that constraints do not rely on neutrino spectra. The intrinsic spectra of astrophysical sources can be uncertain and model dependent, in which the constraints from the spectral modification may suffer from unknown astrophysical systematics. For example, the source spectra are not simple power laws in many models \cite{Murase:2013ffa,Murase:2019vdl}. Furthermore, in the optically thin regime of interactions, the number of interactions between the source and the detector is less than one, in which the spectral modification is expected to be small. On the other hand, if one has a large number of signal events, the echo method allows us to probe the regime of $\tau_\nu<1$ \cite{Murase:2019xqi,Carpio:2022lqk}, which is different from limits set by $\tau_\nu \sim 1$ via the spectral modification \cite{Kelly:2018tyg}. 

Stronger probes of secret neutrino interactions can transform our understanding of the dark sector. An ongoing annihilation of weakly interacting massive particles so-called WIMPs to SM particles indicates possible interaction of DM and SM particles, which is the basis for direct DM detection experiments. There is a distinct possibility that the main portal from the SM to DM might be solely provided by neutrinos, see e.g., Ref.~\cite{Blennow:2019fhy}. ``Scotogenic'' models that generate neutrino mass through its interaction with the dark especially utilize this connection from SM to the dark sector ~\cite{Boehm:2006mi,Farzan:2012sa,Escudero:2016tzx,Escudero:2016ksa,Hagedorn:2018spx,Alvey:2019jzx,Patel:2019zky, Baumholzer19}. Interestingly, the heavy neutrino state proposed by these models can also describe the MiniBooNE anomaly~\cite{Bertuzzo:2018itn,Ballett:2018ynz,Ballett:2019cqp,Ballett:2019pyw}. Moreover, models suggesting an interaction of DM with a heavy sterile neutrino state lead to a direct annihilation of DM to neutrinos ~\cite{Profumo:2017obk,Ballett:2019pyw}. Provided that there exists a sizeable mixing with lighter neutrinos, DM-neutrino interaction would offer the best probe of such interaction models. For more discussion, see ~\cite{Blennow:2019fhy}. 
We should note that potential time delay induced by violation of fundamental symmetries were already discussed in Refs.~\cite{Wang:2016lne,Ellis:2018ogq,Boran:2018ypz,Laha:2018hsh,Wei:2018ajw}, with an emphasis on recent evidence for the coincidence of a high-energy neutrino with a flaring blazar TXS 0506+056. In these scenarios, the features are magnified as the neutrino energy increases. 
However, time delays induced by BSM interactions of neutrinos would be generally inversely proportional to the energy of cosmic neutrinos. As such, BSM searches with astrophysical neutrinos may gain sensitivity thanks to a generally higher level of flux at lower energies. 

In summary, pinpointing sources of astrophysical neutrinos offers a unique and powerful opportunity to search for physics beyond the Standard Model. This work presented the general formalism that is applicable to any scenario that involves a single-scattering process via nonstandard neutrino interaction. Besides the two scenarios discussed here, boosted DM in transient sources of high-energy neutrinos, such as blazars, are another phenomena that the formalism presented in this work can be applied to. The analytical results are useful for precision tests for numerical simulations~\cite{Carpio:2022lqk}. These help us scan wide parameter space, which is often difficult with numerical simulations. This formulation is even applicable to MeV energies targeting neutrino emission from supernovae~\cite{Carpio:2022sml}. As in the case of SN 1987A, supernovae has been already shown to be strong probes of secret neutrino interactions \cite{Kolb:1987qy} (see \cite{Berryman:2022hds} for more details). The echo technique presented in this work offers additional tools for testing secret neutrino interactions.

\begin{acknowledgements}
We thank Jose Carpio for useful discussions. 
The work of K.M. is supported by the NSF Grant No.~AST-1908689, No.~AST-2108466 and No.~AST-2108467, and KAKENHI No.~20H01901 and No.~20H05852. 
A.K. acknowledges the support from Institute for Gravitation and the Cosmos through the IGC postdoctoral fellowship.
\end{acknowledgements}

\bibliography{kmurase,nu}

\end{document}